\documentclass[12pt]{article}
\usepackage[utf8]{inputenc}

\usepackage{tikz,pgfplots}
\usetikzlibrary{arrows.meta,decorations.pathmorphing,
     decorations.markings,shadows,calc}

\pgfplotsset{compat=newest,
      width=0.9\textwidth,height=0.9\textwidth/1.618,
      every tick/.append style={black,line width=1pt},
      every axis/.append style={line width=1pt},
      enlargelimits=false,
      axis lines=middle,
      every inner x axis line/.append style={->},
      every inner y axis line/.append style={->},
      every axis y label/.style={at={(0,1)},above right},
      every axis x label/.style={at={(1,0)},above right},
      every pin edge/.style={solid,black}
}

\providecommand*{\mrm}[1]{\mathrm{#1}}
\providecommand*{\eu}{\ensuremath{\mrm{e}}}
\providecommand*{\iu}{\ensuremath{\mrm{i}}}

\providecommand{\renewoperator}[3]{%
       \renewcommand*{#1}{\mathop{#2}#3}}

\renewoperator{\Re}
               {\mathrm{Re}}{\nolimits}
\renewoperator{\Im}
               {\mathrm{Im}}{\nolimits}

\newif\ifgreek
\def\testgreek#1{
  \ifx#1\alpha\greektrue\else
  \ifx#1\beta\greektrue\else
  \ifx#1\gamma\greektrue\else\ifx#1\Gamma\greektrue\else
  \ifx#1\delta\greektrue\else\ifx#1\Delta\greektrue\else
  \ifx#1\epsilon\greektrue\else
  \ifx#1\zeta\greektrue\else
  \ifx#1\eta\greektrue\else
  \ifx#1\theta\greektrue\else\ifx#1\Theta\greektrue\else
  \ifx#1\iota\greektrue\else
  \ifx#1\kappa\greektrue\else
  \ifx#1\lambda\greektrue\else\ifx#1\Lambda\greektrue\else
  \ifx#1\mu\greektrue\else
  \ifx#1\nu\greektrue\else
  \ifx#1\xi\greektrue\else\ifx#1\Xi\greektrue\else
  \ifx#1\pi\greektrue\else\ifx#1\Pi\greektrue\else
  \ifx#1\rho\greektrue\else
  \ifx#1\sigma\greektrue\else\ifx#1\Sigma\greektrue\else
  \ifx#1\tau\greektrue\else
  \ifx#1\upsilon\greektrue\else\ifx#1\Upsilon\greektrue\else
  \ifx#1\phi\greektrue\else\ifx#1\Phi\greektrue\else
  \ifx#1\chi\greektrue\else
  \ifx#1\psi\greektrue\else\ifx#1\Psi\greektrue\else
  \ifx#1\omega\greektrue\else\ifx#1\Omega\greektrue\else
  \ifx#1\varepsilon\greektrue\else
  \ifx#1\vartheta\greektrue\else
  \ifx#1\varrho\greektrue\else
  \ifx#1\varsigma\greektrue\else
  \ifx#1\varphi\greektrue\else
     \greekfalse
  \fi\fi\fi\fi\fi\fi\fi\fi\fi\fi
  \fi\fi\fi\fi\fi\fi\fi\fi\fi\fi
  \fi\fi\fi\fi\fi\fi\fi\fi\fi\fi
  \fi\fi\fi\fi\fi\fi\fi\fi\fi}

\newcommand{\reg}{\mathcal R} 
\newcommand{\numdensity}{{\mathfrak n}}

\newcommand{\pair}{\mathrm g}
\newcommand{\period}{\ell}

\newcommand{\rv}{\vec{r}}

\newcommand{\vv}{\vec{v}}
\newcommand{\pp}{\mathrm{p}}

\usepackage[most]{tcolorbox} 

\newtcolorbox[auto counter]{optionalnote}[2][]{
    parbox=false,
    colbacktitle= white,
    colback=green!5!white,
    colframe=white!45!black,
    coltitle=black,
    enhanced,
    attach boxed title to top left={yshift=-1mm},
    title={\thetcbcounter.~#2}
,#1}

\newtcolorbox{highlight-result}[1][]{
 parbox=false,
 boxrule=0pt,top=0pt,bottom=0pt,
colback=blue!8!white,
enhanced,#1}

\tcbset{colback=blue!8!white, colframe=white!50!white,top=0mm,bottom=0mm,right=0mm,left=0mm}

\definecolor{Amaranth}{rgb}{0.9, 0.17, 0.31}
\definecolor{Almond}{rgb}{0.94, 0.87, 0.8}
\definecolor{Apricot}{rgb}{0.98, 0.81, 0.69}
\definecolor{Fuchsia}{rgb}{0.57, 0.36, 0.51}
\definecolor{Amethyst}{rgb}{0.6, 0.4, 0.8}

\usepackage[T1]{fontenc}
\usepackage[british]{babel}
\usepackage[autostyle]{csquotes}

\usepackage{charter}
\usepackage{microtype} 

\usepackage{amsmath,amssymb,amsthm,latexsym,amsfonts,mathtools}
\usepackage{nicefrac}
\usepackage{esint}
\usepackage[normalem]{ulem}
\usepackage{ulem}
\usepackage{bm}
\renewcommand{\vec}[1]{\boldsymbol#1}
\usepackage{cases}
\usepackage{empheq}
\usepackage{braket}
\usepackage{cancel}

\usepackage{graphicx,color}
\usepackage{tikz,pgfplots}
\pgfplotsset{compat=1.18} 
\usepackage{psfrag}

\usepackage{float}
\usepackage{subcaption}
\usepackage[font=small,skip=0pt]{caption}

\usepackage{algorithm2e}
\RestyleAlgo{ruled}
\SetKwComment{Comment}{/* }{ */}

\usepackage[left=20mm, right=20mm, top=25mm, bottom=25mm, headheight=25pt, headsep=25pt]{geometry}

\usepackage[colorlinks,bookmarks,pdfusetitle,urlcolor=blue,citecolor=blue,linkcolor=black,bookmarksnumbered,plainpages=false,hyperfootnotes=false]{hyperref}
\usepackage{cleveref}
\usepackage{footmisc}

\usepackage[
  autocite    = superscript,
  backend     = bibtex,
  sortcites   = true,
  style       = numeric,
  doi=false, url=true, isbn=false,
  maxcitenames=1, maxbibnames=1
  ]{biblatex}

\usepackage{fancyhdr}
\usepackage{titlesec}
\titlespacing*{\section}{0pt}{0.4cm}{0.4cm}
\titlespacing*{\subsection}{0pt}{0.4cm}{0.4cm}
\titlespacing*{\subsubsection}{0pt}{0.4cm}{0.4cm}
\titlespacing*{\paragraph}{20pt}{0.0cm}{0.3cm}

\usepackage{verbatim}
\usepackage{marginnote}
\usepackage{enumerate}
\usepackage[textsize=normalsize,textwidth=2cm,colorinlistoftodos]{todonotes}
\usepackage{ctable}
\usepackage{setspace}
\usepackage{amsopn}
\usepackage{epstopdf}
\usepackage{parskip}
\usepackage{etoolbox}
\pretocmd{\chapter}{}{}{}

\appto{\bibsetup}{\sloppy}

\usepackage{authblk}

\makeatletter
\newcommand{\chapterauthor}[1]{
  {\parindent0pt\vspace*{-25pt}
  \linespread{1.1}\large\scshape#1
  \par\nobreak\vspace*{35pt}}
  \@afterheading
}
\makeatother

\begin{document}

\setlength{\parskip}{0pt}
\onehalfspacing

\cleardoublepage
\pagenumbering{roman}

\tableofcontents

\cleardoublepage
\pagenumbering{arabic}

\begin{refsection}

    \section*{\centering \Large \scshape Calculating pair-correlations from random particle configurations} 
    \label{chapter:inter-particle correlations}
    
    \vspace*{15mm}
    
    \chapterauthor{\centering Aris Karnezis, Art L. Gower}{}
    
    \begin{center}
      \large \scshape January 2024
    \end{center}
    
    \vspace*{10mm}
    
    \hrule 

    \vspace*{10mm}

    \section*{\Large \scshape Abstract}

    Particle pair-correlations are broadly used to describe particle distributions in chemistry, physics, and material science. Many theoretical methods require the pair-correlation to predict material properties such as fluid flow, thermal properties, or wave propagation. In all these applications it is either important to calculate a pair-correlation from specific particle configurations, or vice-versa: determine the likely particle configurations from a pair-correlation which is needed to fabricate a particulate material. Most available methods to calculate the pair-correlation from a particle configuration require that the configuration be very large to avoid effects from the boundary. Here we show how to avoid boundary effects even for small particle configurations. Having small particle configurations leads to far more efficient numerical methods. We also demonstrate how to use techniques from smooth nonlinear optimisation to quickly recover a particle configuration from a pair-correlation.

    \newpage

    \section{Introduction}

    \paragraph{Background.} 
    
    In the field of material science, the investigation of structural properties has witnessed significant progress. Molecular dynamics simulations, coupled with neutron and X-ray scattering experiments, have provided valuable insights into the structural characteristics of materials. For disordered materials, most techniques to probe and analyse these structures focus on the pair-correlation function and the structure factor \cite{Pusztai1996, Pusztai1997, McGreevy1988, McGreevy2001, Stillinger2004}. The pair-correlation appears naturally in theoretical methods that use ensemble averaging \cite{Carvalho2020, Salvatore2002, Pathria2011, Carminati2021}, whereas the structure factor appears naturally from scattering experiments \cite{Carvalho2020, Crawford2003}, including small-angle neutron scattering \cite{Sjberg1994, Carsughi1997}. 
    
    The pair-correlation function, denoted by $\pair(r)$, represents the probability of finding a pair of particles separated by a certain distance $r$ \cite{Carvalho2020, Salvatore2002, Pathria2011, Carminati2021}. The structure factor, denoted by $\mathrm{S}(k)$, corresponds to the Fourier transform of $\pair(r)$, and it is essential for characterising the structure of the material from the scattering intensity \cite{Carvalho2020, Crawford2003, Dhabal2017, Piekarewicz2012}.

    Beyond material science, pair-correlations and structure factors extend to calculating shear viscosity, electrical conductivity, and thermal conductivity from a molecular perspective \cite{Nandi2018}. They also play a significant role in exploring complex phenomena such as the \textit{nuclear pasta} observed in the extreme conditions of neutron star crusts \cite{Piekarewicz2012, Horowitz2004, Horowitz2004b}. This phenomenon, characterised by non-uniform arrangements of subatomic particles under extreme gravitational fields, is investigated through neutron and X-ray scattering to understand the internal structure of neutron stars \cite{Piekarewicz2012, Horowitz2004, Horowitz2004b}.

    Knowledge of the pair-correlation function and the structure factor is very useful when designing materials with specific behaviour. Indeed, they can be set in order to achieve specific properties such as hyperuniform structures, materials with exotic band gap profiles and negative refractive index \cite{Torquato2018, Garcia2021, Milosevic2019, Leseur2016, Sheremet2020}.
    By adjusting pair-correlations, we aim to create materials with specific properties and unique structural characteristics.

    From the authors background, the pair-correlation appears when taking an ensemble average of waves in disordered materials \cite{gower2019multiple, gower2021effective, Carminati2021, Tsang2001}. Essentially, the pair-correlation function captures the spatial distribution and arrangement of particles within a material, directly influencing how waves interact with and propagate through the material. That is, the pair-correlation is the only way the material structure affects wave propagation. This suggests a route to design materials to control waves: 
    
    \begin{enumerate}
      \item To determine pair-correlations that lead to band-gaps, frequency filters, or enhanced transmission.
      \item To determine configurations of particles that match the desired pair-correlation.
    \end{enumerate}

    Step 2) is known as the realizability problem \cite{Zhang2020,Kuna2007, Costin2004, Crawford2003,Uche2006}, and there are open questions about when it is possible to solve \cite{Pothoczki2010, Fbin2007, Fujii2007, Cormier1997, Keen1990, McGreevy1988}. 
    
    \paragraph{Realizability problem.} 
    
    Determining a likely configuration of particles from a specific pair-correlation is called the realizability problem. To make the problem easier to understand, here is an example: Think about solving a jigsaw puzzle. Each puzzle piece is like a particle in a material. The goal is to fit all these pieces together within the puzzle board to create a picture. This picture represents our target arrangement of particles. However, just like puzzle pieces must fit together in a certain way and cannot be forced into the wrong place or overlap, the particles in our material must also be arranged in a physically feasible way. This means we can not just place them in any random configuration; they need to fit together according to physical laws, just like puzzle pieces. So, realizability is about making sure that the arrangement we come up with can actually exist in the real world.
    
    The realizability problem appears in the study of many-body systems, such as liquids, and disordered materials \cite{Stillinger2004, Uche2006, Kuna2007, Zhang2020, Kuna2011, Costin2004, Torquato2002, Crawford2003, Yamada1961}. One of the challenges is that the problem often lacks a unique solution. Multiple configurations of particles can result in the same pair-correlation, especially in disordered media where the properties and behaviour of particulates can be influenced by factors like particle size, shape and density \cite{Crawford2003, Torquato2006}. 
    
    Several necessary conditions for potential pair-correlation functions have been identified, including the requirement of non-negativity to ensure realistic representation of properties such as density or probability distributions in materials, and restrictions on their associated structure factors, which measure variations in particle density within a material \cite{Stillinger2004}. However, establishing a set of sufficient conditions for these functions remains an open challenge \cite{Stillinger2004, Salvatore2002}. We note that the problem is not completely resolved even in just one spatial dimension.
    
    \paragraph{Reverse Monte-Carlo.} 
    
    Reverse Monte-Carlo structural modelling is one technique that has been used to calculate particle configurations that match a measured structure factors or pair-correlations \cite{Pothoczki2010, Fbin2007, Fujii2007, Cormier1997, Keen1990, McGreevy1988,Torrie1977}. Typically these Monte-Carlo simulations are guided by a Genetic Algorithm, or similar random searches, which are computationally intensive \cite{McGreevy1988, McGreevy2001, Manwart1999, Quiblier1984, Torrie1977}. These methods are brute-force, and typically use non-gradient-based methods. While effective to some extent, they struggle when dealing with pair-correlation functions that lack smoothness. In response to this limitation, we propose a novel approach based on smooth optimisation. Given the straightforward computations of gradients, it stands to reason that gradient-based optimisation methods \cite{ruszczynski2011nonlinear} hold significant potential for superior performance over traditional non-gradient-based approaches. This advantage comes from their ability to efficiently move through the search space by using directional information, thus offering a more targeted and faster convergence to optimal particle arrangements.

    \paragraph{Paper summary.}

    In \Cref{sec:particle distributions}, we start by deducing in a simple self-contained way how to describe probability distributions and the pair-correlation in terms of a set of particles. In \Cref{sec:two-regions} we show how to avoid the effects of boundaries when calculating pair-correlations. The effects of boundaries are usually undesirable, and the details we present seem to be missing from most references. In \Cref{sec:discrete isotropic form} we show the same calculations again, but without the use of Dirac deltas for didactic purposes. In \Cref{sec:structure factor} we deduce the structure factor for isotropic distributions both from any given pair-correlation and a set of particles. These results are needed in \Cref{sec:particles from structure}, where we present a method to calculate a configuration of particles that matches a given structure factor. Developing more efficient methods to reconstruct particle configurations from pair-correlations remains an ongoing challenge in material science and computational chemistry \cite{Patelli2009}. The method we propose uses techniques from smooth nonlinear optimisation to improve the efficiency, which we are able to do because the structure factor is a smooth function of the particle positions. We also present some preliminary numerical results. Finally, in \Cref{sec:discussion} we summarise what the paper achieved and possible future directions.
        
    \section{Particle distributions} \label{sec:particle distributions}
    
    Consider there are $J$ particles placed within some region $\reg$. That is, the centre of every particle $\vec r_j \in \reg$.  We represent one possible configuration of particles, or ensemble, by the set $\mathcal X^s$, where every $\rv \in \mathcal X^s$ is the centre of a particle in the ensemble.
    
    The function $\mathrm p(\rv)$ is the probability density of finding a particle centred at $\rv$. We can approximate $\mathrm p(\rv)$ by defining a mesh of volume elements $V(\vv_i)$, where the vector $\vv_i$ is the centre of the volume element, and then counting the number of particles in each $V(\vv_i)$ divided by the total number of particles. This allows us to introduce $\mathrm p_V(\vv_i)$ as an approximation to $\mathrm p(\vv_i)$, expressed as $\mathrm p(\vv_i) \approx \mathrm p_V(\vv_i)$, to estimate the probability:
    \begin{equation}\label{eqn:approx-prob-density}
      \mathrm p_V(\vv_i) = \frac{1}{S}\sum_{s}^S \frac{\# \left[ \mathcal X^s \cap V (\vv_i) \right]}{\# \mathcal X^s} \frac{1}{|V (\vv_i)|},
    \end{equation}
    where $|V(\vv_i)|$ is the volume of $V(\vv_i)$, $\# \mathcal X$ is the number of elements in the set $\mathcal X$, and $S$ is the number of ensembles considered. The use of a finite number of ensembles introduces a degree of statistical uncertainty. While increasing $S$ can improve the accuracy of the approximation, it also increases computational demands. There is a trade-off between computational efficiency and statistical robustness. The chosen value of $S$ should provide a reliable representation of particle distributions while remaining computationally manageable.
    
    \noindent We require that the mesh:
    \[
      V(\vv_i) \cap V(\vv_j) = \varnothing, \;\; \text{for} \;\; j \not = i,  
    \]
    ensuring that each volume element is distinct and non-overlapping. This is crucial for the independence of the measurements. It prevents double-counting of particles and ensures that each volume element contributes uniquely to the probability density calculation.

    In a similar {way}, we can approximate $\mathrm p(\vec x_1, \vec x_2)$, which is the joint probability density of finding one particle centred at $\vec x_1$ and another centred at $\vec x_2$, while averaging over all over particle positions. If we assume that both $\vec x_1$ and $\vec x_2$ are distributed within $\reg$, then we can approximate the function $\mathrm p(\vec x_1, \vec x_2)$ with the formula:
    \begin{equation}\label{eqn:approx-pair-corr}
    \mathrm p_V(\vv_i, \vv_j) = \frac{1}{S} \sum_{s}^S\frac{\# \left[\mathcal X^s \cap V (\vv_i)\right]}{\# \mathcal X^s} \frac{\# \left[\mathcal X^s \cap V (\vv_j)\right]}{\# \left[\mathcal X^s \backslash V (\vv_i) \right]} \frac{1}{|V (\vv_i)| |V (\vv_j)|}, \quad \text{for} \;\; i \not = j,
    \end{equation}
    where $\mathrm p_V(\vv_i, \vv_i) = 0$ for every $i$, and $\mathcal X^s \backslash V (\vv_i)$ is defined as the set $\mathcal X^s$ without $V (\vv_i)$. Note that:
    \[
      \mathrm p(\vv_i, \vv_j) \approx \mathrm p_V(\vv_i, \vv_j),
    \] 
    for every $\vv_i$ and $\vv_j$. In simple terms, it allows us to estimate the probability of finding pairs of particles at specific positions within a material by statistically approximating the probability.

    As expected, using this approximation, the integral of $\mathrm p(\vec x_1, \vec x_2)$ for $\vec x_1, \vec x_2 \in \reg$ gives one: 
    \begin{align} \label{eqn:check-joint-discrete}
      \int \mathrm p(\vec x_1, {\vec x_2}) \mathrm d \vec x_1 \mathrm d \vec x_2 &\approx \sum_{i,j} \mathrm p_V(\vv_i, \vv_j) |V(\vv_i)| |V(\vv_j)| \nonumber \\
      &=\frac{1}{S} \sum_{s}^S \sum_{i}\sum_{j \not = i} \frac{\# \left[\mathcal X^s \cap V (\vv_i)\right]}{\# \mathcal X^s} \frac{\# \left[\mathcal X^s \cap V (\vv_j)\right]}{\# \left[\mathcal X^s \backslash V (\vv_i) \right]} \nonumber \\
      &= \frac{1}{S} \sum_{s}^S \sum_i \frac{\# \left[\mathcal X^s \cap V (\vv_i)\right]}{\# \mathcal X^s} \frac{\# \left[\mathcal X^s \cap  \cup_{j \not = i} V (\vv_j)\right]}{\# \left[\mathcal X^s \backslash V (\vv_i) \right]} \nonumber \\
      &= \frac{1}{S} \sum_{s}^S \frac{\# \left[\mathcal X^s \cap \cup_i V (\vv_i)\right]}{\# \mathcal X^s} = 1,
      \end{align}      
    where we used that $\cup_i V (\vv_i) = \reg$, $\mathrm d \vec x_1  \approx |V (\vv_i)|$,$\mathrm d \vec x_2 \approx |V (\vv_j)|$, and $\reg \cap  \cup_{j \not = i} V (\vv_j) = \reg \backslash V(\vv_i)$ which implies that $\mathcal X^s \cap  \cup_{j \not = i} V (\vv_j) = \mathcal X^s \backslash V(\vv_i)$.
    
    The term that often appears in methods that use ensemble averaging of particulates \cite{Bringi1982,Tsang2001,Salvatore2002} is the particle pair-correlation, which is defined as:
    \begin{equation}\label{def:pair-correlation}
      \pair(\vec x_1, \vec x_2) := \frac{J - 1}{J} \frac{\pp(\vec x_1, \vec x_2)}{\mathrm p(\vec x_1) \mathrm p(\vec x_2)},
    \end{equation}
    where $J = \# \mathcal X_s$ is the total number of particle, where we have assumed that every configuration $\# \mathcal X_s$ has the same number of particles. The factor \(\dfrac{J - 1}{J}\) is correction for a finite number of particles in a region when calculating the pair-correlation function \(\pair(\vec{x}_1, \vec{x}_2)\). In an infinite region, as particles become uncorrelated due to distance or lack of interaction, the pair-correlation approaches 1, indicating a random distribution without any correlation between particle positions. However, in a finite region, the presence of a limited number of particles introduces a bias because each particle is slightly more likely to be closer to another particle than it would be in an infinite region. The correction factor \(\dfrac{J - 1}{J}\) adjusts for this bias by scaling down the correlation as the number of particles increases, ensuring that \(\pair(\vec{x}_1, \vec{x}_2) \rightarrow 1 \) for large distances between particles or when particles become uncorrelated as confirmed by \cite[Equation (8.1.2)]{Tsang2001}. We demonstrate this in the next section. 
    
    \subsection{Particles as Dirac delta}

    For a finite, but very large number, of particles we can rewrite the pair-correlation in terms of Dirac deltas. For example, turning to ~\eqref{eqn:approx-prob-density} we assume there is a finite number of ensembles $S$, each with a finite number of particles $J$. Then we can make the volume elements $V (\vv_i)$ small enough so that there is at most one particle in each volume element, which implies that $\# \mathcal X^s = J$ and:
    \begin{equation} \label{eqn:cases_for_cap}
      \# \left[ \mathcal X^s \cap V (\vv_i) \right] =  \begin{cases}
        1, & \text{if there is a particle in} \;\; V (\vv_i),  
        \\
        0, & \text{if there is no particle in} \;\; V (\vv_i).
      \end{cases}
    \end{equation} 
    For notational convenience, we choose the mesh of volume elements such that each element is centred at a particle $\rv_i \in \mathcal X^s$, if it contains a particle. Equation \eqref{eqn:cases_for_cap} allows us to rewrite \eqref{eqn:approx-prob-density}  in the form:
    \begin{equation}
      \mathrm p_V(\rv_i) = \frac{1}{S}\sum_{s}^S \frac{1}{|V (\rv_i)|} \frac{1}{J},
    \end{equation}
    for every $\rv_i \in \mathcal X^s$.
    
    As we are taking the limit of all the volume elements going to zero, $|V(\rv_i)| \to 0$, we can approximate:
    \[
      \mathrm p(\vec x) =  \begin{cases}
        \mathrm p_V(\rv_i), & \quad \text{if} \;\; \vec x \in V(\rv_i),  
        \\
        0, & \quad \text{else}, 
      \end{cases}
    \] 
    where $\mathrm p(\vec x) = 0 $ if $\vec x$ is in a volume element that does not contain a particle. With this definition, and taking the limit $|V (\vv_i)| \to 0$ for every $i$, we can rewrite \eqref{eqn:approx-prob-density} in the form:
    \begin{equation} \label{eqn:p-dirac} 
      \mathrm p(\vec x) = \frac{1}{J S}\sum_{s}^S \sum_{\rv_i \in \mathcal X^s}\delta(\vec x - \rv_i),
    \end{equation}
    where $\delta(\vec x - \rv_i)$ is the Dirac delta function. This allows each volume element to represent an infinitesimally small point in space, effectively turning each particle into a point particle. For details on the Dirac delta see \cite{arfken2011mathematical}.

    Repeating analogous steps for the joint probability density \eqref{eqn:approx-pair-corr} we obtain:
    \begin{equation}\label{eqn:p-joint-dirac}
      \mathrm p(\vec x_1, \vec x_2) = \frac{1}{S} \frac{1}{J(J - 1)} \sum_{s}^S \sum_{\rv_i \in \mathcal X^s} \sum_{\rv_j \not = \rv_i, \rv_j \in \mathcal X^s} \delta(\vec x_1 - \rv_i) \delta(\vec x_2 - \rv_j).
    \end{equation}
    In \eqref{eqn:p-joint-dirac}, the outer summation over all ensembles $S$ accounts for the randomness and is essential for capturing the statistical variation across different possible configurations of particles within the material. The inner summation considers all pairs of particles $\rv_i$ and $\rv_j$ within each ensemble, excluding cases where $\rv_i = \rv_j$ because a particle cannot pair with itself. This guarantees that the model accurately represents the joint probability of finding two distinct particles at two specific points.

    This method provides a robust and flexible framework for statistical analysis across a wide range of densities, especially in disordered systems. However, it can be computationally intensive due to the high level of detail and precision it requires. As each particle is represented as a point in space, the computations involve handling potentially large numbers of Dirac delta functions.

    \subsection{Isotropic distributions} \label{sec:isotropic-L}

    For most of this paper we focus on homogeneous and isotropic distributions. The homogeneous assumption dictates that the particles are equally likely to be placed anywhere in space, which leads to $\pp(\rv)$ being a constant. If we choose all volume elements to have the same volume $|V(\vv_i)| = |V|$, then using that $\pp_V(\vv_i)$ is a constant together with \eqref{eqn:approx-prob-density} leads to:
    \begin{equation} \label{eqn:p-constant}
        \mathrm p_V(\vv_j) = \frac{1}{S I} \sum_{i} \sum_{s}^S \frac{\# \left[ \mathcal X^s \cap V (\vv_i) \right]}{\# \mathcal X^s} \frac{1}{|V|} = \frac{1}{S I} \frac{1}{|V|} \sum_{s}^S \frac{\# \left[ \mathcal X^s \cap \cup_i V (\vv_i) \right]}{\# \mathcal X^s} =  \frac{1}{I|V|} = \frac{1}{|\reg|},   
    \end{equation}
    where $I$ is the total number of volume elements $V(\vv_i)$, $|\reg|$ is the volume of $\reg$, and we used $\cup_i V (\vv_i) = \reg$ and similarity $I |V| = |\reg|$. 

    For an isotropic distribution we have that only the inter-particle distance is needed for their joint probability \cite{Pathria2011, Philcox2023}. That is $\pp(\vec x_1,\vec x_2) = \pp(|\vec x_1 -\vec x_2|)$, where $|\rv|$ is the magnitude of the vector $\rv$. To use this property to simplify the formula for the pair-correlation~\eqref{def:pair-correlation}, we start by writing:
    \begin{equation}\label{eqn:isotopic-property-g}
        \pair(|\vec x_1 -\vec x_2|) = \pair(\vec x_1,\vec x_2),
    \end{equation}
    then integrate both sides over all values such that $|\vec x_1 -\vec x_2| = z$ for fixed $z$ relative to the variables of integration. To make this clearer we introduce the ball region using standard set-builder notation: 
    \begin{equation}
      \mathcal B(\vec x; r) = \{\vec y \in \mathbb R^3: \; |\vec x - \vec y | \leq r \}.
    \end{equation}
    The ball region is introduced to simplify the process of integrating over distances. It defines a spherical region of radius $r$ around a point $\vec x$, which is essential for transforming the pair-correlation function into a function of distance alone.

    Using this notation, we will integrate both sides of \eqref{eqn:isotopic-property-g} over every $\vec x_2 \in \partial \mathcal B(\vec x_1; z) \cap \mathcal{R}$, followed by over all $\vec x_1 \in \reg$, which results in:
    \begin{equation}\label{eqn:isotopic-property-g-2}
      \int_{\reg} \int_{\partial \mathcal B(\vec x_1; z) \cap \reg } \pair(|\vec x_1 -\vec x_2|)  \mathrm d S_2 \mathrm d \vec x_1 =   \int_{\reg} \int_{\partial \mathcal B(\vec x_1; z) \cap \reg }\pair(\vec x_1,\vec x_2)  \mathrm d S_2 \mathrm d \vec x_1.
    \end{equation}
    \noindent Here, $\vec x_1$ represents the position of a particle within the region $\mathcal{R}$, $\vec x_2$ is the position of another particle, and $\partial \mathcal B(\vec x_1; z)$ is the surface of a sphere with radius $z$ centred at $\vec x_1$. Physically, this means we are looking at all possible positions a second particle could occupy that are exactly a distance $z$ away from the first particle, and we are integrating the pair-correlation function over the spherical surface that lies within the region $\mathcal{R}$, to understand how particle density varies at this specific separation distance.
    
    \noindent The left side of \eqref{eqn:isotopic-property-g-2} can move out of the integral, resulting in:
    \begin{equation} \label{eqn:L-integral}
      \int_{\reg} \int_{\partial \mathcal B(\vec x_1; z) \cap \reg } \pair(z) \mathrm d S_2 \mathrm d \vec x_1   = L(z) \pair(z), \;\; \text{with} \;\; L(z) = \int_{\reg} \int_{\partial \mathcal B(\vec x_1; z) \cap \reg }  \mathrm d S_2 \mathrm d \vec x_1.
    \end{equation}
    For the right side of \eqref{eqn:isotopic-property-g-2} we use the definition~\eqref{def:pair-correlation}, and then rearrange \eqref{eqn:isotopic-property-g-2} to reach:
    \begin{align}\label{eqn:reduce-pair-corr}
      & \pair(z) = \frac{J - 1}{J}\frac{1}{L(z)} \int_{\reg} \int_{\partial \mathcal B(\vec x_1; z) \cap \reg} \frac{\mathrm p(\vec x_1, \vec x_2)}{\mathrm p(\vec x_1)\mathrm p(\vec x_2)} \mathrm d \vec x_1 \mathrm d S_2.
    \end{align}
    To reach a simple formula to calculate $\pair(z)$, we need to calculate the integral $L(z)$. 
    In general, calculating $L(z)$ can be awkward and we show how to avoid this in the next section. 

    
    To simplify the radial pair-correlation~\eqref{eqn:reduce-pair-corr} we combine \eqref{eqn:p-joint-dirac} and \eqref{eqn:p-constant} to write the pair-correlation \eqref{def:pair-correlation} in terms of Dirac delta functions, which we then substituted into \eqref{eqn:reduce-pair-corr}, and integrate over $\vec x_1$ to obtain:
    \begin{align}\label{eqn:dirac-radial-1}
       \pair(z) 
       & = \frac{1}{S} \frac{1}{\numdensity^2} \sum_{s}^S \sum_{\rv_i \in \mathcal X^s} \sum_{\rv_j \not = \rv_i, \rv_j \in \mathcal X^s}  \frac{1}{L(z) } \int_{\partial \mathcal B(\vec r_i; z) \cap \reg}  \delta(\vec x_2  - \rv_j)  \mathrm d S_2.
    \end{align}
    where $\numdensity : = J / |\mathcal{R}|$ is the particle number density.
    To simplify \eqref{eqn:dirac-radial-1}, note that because $\rv_j \in \reg$ we have that:
    \[
      \int_{\partial \mathcal B(\vec r_i; z) \cap \reg}  \delta(\vec x_2  - \rv_j)  \mathrm d S_2 = \int_{\partial \mathcal B(\vec r_i; z)}  \delta(\vec x_2  - \rv_j)  \mathrm d S_2.  
    \]
    At greater length, this is a result of the integral over $\vec x_2 \not \in \reg$ having an integrand which is zero, $\delta(\vec x_2  - \rv_j) =0$, because $\rv_j \in \reg$, allowing us to rewrite \eqref{eqn:dirac-radial-1} in the form: 
    \begin{align}\label{eqn:dirac-radial-2}
       \pair(z) 
       & = \frac{1}{S} \frac{1}{\numdensity^2} \sum_{s}^S \sum_{\rv_i \in \mathcal X^s} \sum_{\rv_j \not = \rv_i, \rv_j \in \mathcal X^s}  \frac{1}{L(z) } \int_{\partial \mathcal B(\vec 0; z)}  \delta(\vec r_i  - \rv_j+ \vec x )  \mathrm d S_x,
    \end{align}
    where we used the change of variables $\vec x = \vec x_2 - \rv_i$. The change of variables to $\vec x = \vec x_2 - \rv_i$ essentially shifts the centre of the sphere to the origin. We can now show how the Dirac delta function can be used to isolate the contribution of particle pairs at exactly distance $z$ apart:
    \begin{equation}\label{eqn:delta 3d to z}
      \int_{\partial \mathcal B(\vec 0; z)}  \delta(\vec r + \vec x ) \mathrm d S_x = \delta(|\vec r| - z),
    \end{equation}
    by noting that for any $z_1 < z_2$ we have that:
    \begin{equation} \label{eqn:cases_for_dirac}
      \int_{z_1}^{z_2} \int_{\partial \mathcal B(\vec 0; z)}  \delta(\rv + \vec x )   \mathrm d S_x \mathrm d z = 
      \int_{\mathcal B(\vec 0; z_2 ) \setminus \mathcal B(\vec 0; z_1)}  \delta(\rv + \vec x )   \mathrm d V_x  =
      \begin{cases}
        1, & \text{if} \;\; z_1 < |\rv| < z_2,
        \\
        0, & \text{else},
      \end{cases}
    \end{equation}
    where we note that $\mathrm d S_x \mathrm d z = \mathrm d V_x$ is a volume element. Considering the properties of the Dirac delta \cite{arfken2011mathematical}, equation \eqref{eqn:cases_for_dirac} can be used to deduce \eqref{eqn:delta 3d to z}. 
    
    Substituting these results into \eqref{eqn:dirac-radial-2} leads to:
    \begin{align}\label{eqn:dirac-radial}
      \pair(z) 
      & = \frac{1}{S} \frac{1}{\numdensity^2 L(z) } \sum_{s}^S \sum_{\rv_i \in \mathcal X^s} \sum_{\rv_j \not = \rv_i, \rv_j \in \mathcal X^s}    \delta(|\rv_i - \rv_j| - z).
    \end{align}
    The only change in \eqref{eqn:dirac-radial} when changing from three spatial dimensions to two spatial dimensions is that $L(z)$ is given by~\eqref{eqn:L-integral} but with $\reg$ being two dimensional and the integral over $S_2$ being a line integral.
    
    \begin{figure}[ht]
      \centering
      \includegraphics[width=0.492\linewidth]{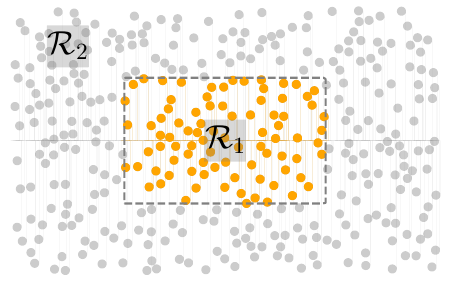}
      \includegraphics[width=0.492\linewidth]{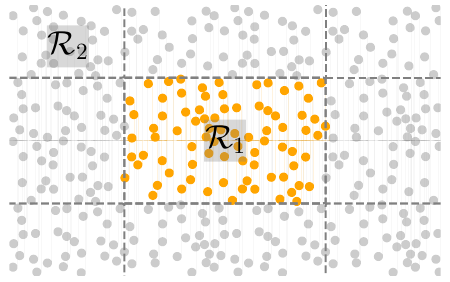}
      \caption{The left image shows a finite set of particles in a region $\reg_1$ taken from a larger set of disordered particles in the region $\reg_2$. Note that $\reg_1$ is contained within $\reg_2$. On the right a unit cell of random particles in a region $\reg_1$ that is periodically tilled. On the right, the region $\reg_2$ is a cut out from the periodic tilling of the particles in $\reg_1$. We use $\period$ to indicate the minimum length of periodicity, which is the height of the unit cell shown in the image on the right.}
      \label{fig:finite-sets}
      \end{figure}
    
    \section{Particles in two regions} \label{sec:two-regions}

    To avoid the influence of the boundary that encloses the particles, and avoid calculating $L(z)$ appearing in \eqref{eqn:reduce-pair-corr-2}, we consider two different regions where particles can be placed $\reg_1$ and $\reg_2$ as shown in \Cref{fig:finite-sets}. We consider that $\vec x_1$ and $\vec x_2$ are two different random variables with $\vec x_1 \in \reg_1$ and $\vec x_2 \in \reg_2$. Note that $\reg_1 \subset \reg_2$, so, the particles $J_1$ of $\reg_1$ are a subset of particles $J_2$ of $\reg_2$. This setup allows for a focus on the internal interactions within $\reg_1$ while minimising the boundary effects from $\reg_2$.

    To calculate the joint probability distribution $\pp(\vec x_1,\vec x_2)$, let $\mathcal X^s$ be such that every $\vec x_2 \in \mathcal X^s$ is also in $\vec x_2 \in \reg_2$, then considering that $\vec x_1 \in \reg_1$ and $\vec x_2 \in \reg_2$ we have that:
    \begin{equation}\label{eqn:approx-pair-corr-2}
    \mathrm p_V(\vv_i, \vv_j) = \frac{1}{S} \sum_{s}^S\frac{\# \left[\mathcal X^s \cap V (\vv_i)\right]}{\# [\mathcal X^s \cap \reg_1]} \frac{\# \left[\mathcal X^s \cap V (\vv_j)\right]}{\# [\mathcal X^s \backslash V(\vv_i)]}
    \frac{1}{|V (\vv_i)| |V (\vv_j)|} \quad \text{for} \;\; i \not = j,
    \end{equation}
     such that $\vv_i \in \reg_1$, $\vv_j \in \reg_2$, and $V(\vv_i) \cap V(\vv_j) = \varnothing$ for every $i,j$, and:
     \[
      \cup_{i=1}^{J_1} V (\vv_i) = \reg_1 \quad \text{and}  \quad \cup_{j=1}^{J_2} V (\vv_j) = \reg_2.
      \]
     As a check of the formula \eqref{eqn:approx-pair-corr-2}, performing an integral of $\mathrm p_V(\vv_i, \vv_j)$ for $\vv_i \in \reg_1$ and $\vv_j \in \reg_2$ gives 1 as expected by following similar steps as shown in~\eqref{eqn:check-joint-discrete}.
    
    For the two different regions, the definition of the particle pair-correlation~\eqref{def:pair-correlation} now becomes:
    \begin{equation}\label{def:pair-correlation-two}
      \pair(\vec x_1, \vec x_2) := \frac{J_2 - 1}{J_2} \frac{\pp(\vec x_1, \vec x_2)}{\mathrm p(\vec x_1) \mathrm p(\vec x_2)},
    \end{equation}
    where $J_2 = \# \mathcal X^s$. 
    
    \subsection{Isotropic distributions}
    
    The probability density for the two regions, and an isotropic distribution of particles, is similar to before with:
    \begin{equation} \label{p:two-regions}
      \mathrm{p}(\vec x_1) = \frac{1}{|\reg_1|} \quad \text{and} \quad \mathrm{p}(\vec x_2) = \frac{1}{|\reg_2|}.
    \end{equation}
    Using analogous steps shown in \Cref{sec:isotropic-L} to reach \eqref{eqn:dirac-radial} we can reach:
    \begin{equation} \label{eqn:isotropic-L-delta-2}
      \pair(z) = \frac{1}{\numdensity^2  L(z) }\frac{1}{ S} \sum_{s}^S \sum_{r_i \in \mathcal X^s \cap \reg_1} \sum_{r_j \not = r_i, r_j \in \mathcal X^s} \delta(|\rv_i - \rv_j| - z).
    \end{equation}
    Equation \eqref{eqn:isotropic-L-delta-2} "selects" only those particle pairs whose separation distance is exactly $z$. It does this by being zero everywhere except where its argument is zero. In this case, it becomes non-zero only when $|\rv_i - \rv_j| = z$.

    However, now we can explicitly calculate $L(z)$ given by:
    \begin{equation}\label{eqn:L-2}
      L(z) = \int_{\reg_1} \int_{\partial \mathcal B(\vec x_1; z) \cap \reg_2}  \mathrm d S_2 \mathrm d \vec x_1.
    \end{equation}
    Assume we want to calculate $\pair(z)$ for $0 \leq z \leq Z$. To do this we can require that: the distance between the boundaries $\partial \reg_2$ and $\partial \reg_1$ be greater than or equal to $Z$. This condition ensures that for any $z \leq Z$, any point $ \vec x_2$ on the surface of the sphere $\mathcal B(\vec x_1; z)$ is also within $\reg_2$. This is crucial for the validity of the pair-correlation calculation, as it guarantees that the integration domain over $S_2$ becomes the entire surface of the sphere, simplifying the computation of $L(z)$. In other words, the domain of integration over $S_2$ becomes $\partial \mathcal B(\vec x_1; z) \cap \reg_2 = \partial \mathcal B(\vec x_1; z)$, which used in \eqref{eqn:L-2} for three spatial dimensions leads to:
    \begin{equation}\label{eqn:L-sphere}
      L(z) = 4\pi z^2 |\reg_1|.
    \end{equation}
    As a quick check, if $\reg_2$ was a sphere with radius $R_2$ centred at the origin, then, 
    $
      \int_0^{R_2} L(z) \mathrm d z 
      = |\reg_1| |\reg_2| 
    $
    as it should. Substituting \eqref{eqn:L-2} into \eqref{eqn:isotropic-L-delta-2} leads to:
    \begin{equation} \label{eqn:isotropic-L-delta-3D}
        \tcboxmath{
        \pair_3(z) = \frac{1}{ 4\pi \numdensity z^2 J_1}\frac{1}{ S} \sum_{s}^S \sum_{r_i \in \mathcal X^s \cap \reg_1} \sum_{r_j \not = r_i, r_j \in \mathcal X^s} \delta(|\rv_i - \rv_j| - z),  \quad \text{(3D spatial),}}
    \end{equation}
    where $J_1 = \# [\mathcal X^s \cap \reg_1]$, and we use $\pair_3$ instead of $\pair$ to indicate that \eqref{eqn:isotropic-L-delta-3D} is for 3 spatial dimensions.
    
    For two spatial dimensions we have $L(z) = 2 \pi z |\reg_1|$ which substituted into \eqref{eqn:isotropic-L-delta-2} leads to:
    \begin{equation} \label{eqn:isotropic-L-delta-2D}
    \tcboxmath{
      \pair_2(z) = \frac{1}{2 \pi \numdensity z J_1  }\frac{1}{ S} \sum_{s}^S \sum_{r_i \in \mathcal X^s \cap \reg_1} \sum_{r_j \not = r_i, r_j \in \mathcal X^s} \delta(|\rv_i - \rv_j| - z), \quad \text{(2D spatial).}}
    \end{equation}

    Next, we re-deduce \eqref{eqn:isotropic-L-delta-3D} and \eqref{eqn:isotropic-L-delta-2D} without using Dirac delta functions. 
    
    \section{The discrete form for isotropic pair-correlations} \label{sec:discrete isotropic form}
    
    In this section we redo the calculations that lead to \eqref{eqn:isotropic-L-delta-3D} and \eqref{eqn:isotropic-L-delta-2D} but without the use of the Dirac delta function. That is, without taking the limit of volume elements tending to zero as used to reach \eqref{eqn:p-dirac}. Doing this serves two purposes: 
    \begin{enumerate}
      \item It can be simpler to understand these discrete formulas, and implement them as a numerical method.
      \item It helps to verify the formulas by reaching formulas we can compare with literature and having two avenues to deduce the same formulas.
    \end{enumerate}

    \subsection{Particles in one region}
    
    Here we consider that all particles are within one region, $\rv_i, \rv_j \in \mathcal X^s$, and re-deduce the results that led to the formula \eqref{eqn:dirac-radial} but without Dirac deltas.  
    
    To reach a simple formula for $\pair(z)$ with the discrete approximation \eqref{eqn:approx-pair-corr}, we discretise the integral in~\eqref{eqn:reduce-pair-corr}, substitute~\eqref{eqn:approx-pair-corr}, and use the discrete approximation for the differentials:
    \begin{equation} \label{eqn:discrete-volume-elements}
      \mathrm d \vec x_1 = |V(\vv_i)|, \quad  \mathrm d z \mathrm d S_2 =|V(\vv_j)|,
    \end{equation}
    to obtain:
    \begin{equation}\label{eqn:reduce-pair-corr-disc}
      \pair(z)  = \frac{J - 1}{J}\frac{|\reg|^2}{L(z) \mathrm dz S} \sum_{s}^S \sum_i \sum_{|\vv_i - \vv_j| \approx z}
    \frac{\# \left[\mathcal X^s \cap V (\vv_i)\right]}{ \# \mathcal X^s} \frac{\# \left[\mathcal X^s \cap V (\vv_j)\right]}{ \# \left[\mathcal X^s \backslash V (\vv_i)\right]},
    \end{equation}
    where we used~\eqref{eqn:p-constant} to substitute $\mathrm{p}(\vec x_1) = \mathrm{p}(\vec x_2) = 1/|\reg|$, and the sum over $j$ is for every $\vv_j$ such that $z - \mathrm dz/2 < |\vv_i - \vv_j| < z + \mathrm dz/2$. As the minimum value for $z = 2 a$, we just set $\pair(z) = 0$ for $z \leq 2a$, where $a$ denotes the radius of the particle.

    At this point, due to the choice~\eqref{eqn:discrete-volume-elements}, the volumes $|V(\vv_i)|$ and $|V(\vv_j)|$ do not appear explicitly in \eqref{eqn:reduce-pair-corr-disc}.
    This allows us to simplify the formula by choosing $V(\vv_i)$ and $V(\vv_j)$ to be sufficiently small so that they each contain no more than one particle. For the indices $i$ and $j$ where there is no particle in $V (\vv_i)$ and $V (\vv_j)$ respectively, we have $\# \left[\mathcal X^s \cap V (\vv_i)\right] =0$ and $\# \left[\mathcal X^s \cap V (\vv_j)\right] =0$. This makes it convenient now to only sum over the $i$, and $j$ where:
    \begin{align}
    & \# \left[\mathcal X^s \cap V (\vv_i)\right] = \# \left[\mathcal X^s \cap V (\vv_j)\right] = 1,
      \\
    & \# \left[\mathcal X^s \backslash V (\vv_i)\right] = \# \mathcal X^s - 1 = J -1,
    \end{align}
    which used in \eqref{eqn:reduce-pair-corr-disc} leads to:
    \begin{equation}\label{eqn:reduce-pair-corr-disc-2}
      \pair(z) = \frac{1}{\numdensity^2}\frac{1}{L(z) dz S} \sum_{s}^S \sum_i \sum_{|\vv_i - \vv_j| \approx z} 1,
    \end{equation}
    where we used the particle number density $\numdensity : = J/|\reg|$. Now, when the volume elements $V(\vv_i)$ and $V(\vv_j)$ are significantly small, approaching zero there is only one particle centre in $V(\vv_i)$ and one other particle centre in $V(\vv_j)$. In this limit, $\vv_i$ and $\vv_j$ approximate the actual positions $\rv_i$ and $\rv_j$ in $\mathcal X^s$, which means we can sum over the actual positions $\rv_i$ and $\rv_j$ instead of the volume elements $\vv_i$ and $\vv_j$. This leads us to rewrite \eqref{eqn:reduce-pair-corr-disc-2} in the reduced form:
    \begin{equation}\label{eqn:reduce-pair-corr-2}
      \pair(z)
      =  \frac{1}{\numdensity^2 L(z) \mathrm dz S} \sum_{s}^S \sum_{\rv_i\in \mathcal X^s} \# \mathcal X^s_i(z), \quad \text{(discrete pair-correlation),}
    \end{equation}
    where $\mathcal X^s_i(z)$ are all the particles $\rv_j$ such that are $|\rv_i - \rv_j| \approx z$, or, more precisely with set builder notation:
    \begin{equation}
      \mathcal X^s_i(z) := \{\rv_j \in \mathcal X^s: z - \mathrm dz /2 \leq |\rv_i - \rv_j| < z + \mathrm dz /2 \},
    \end{equation} 
    and $\# \mathcal X^s_i(z)$ is the number of elements in $\mathcal X^s_i(z)$. 
    
    The only change in the formula \eqref{eqn:reduce-pair-corr-disc-2} when changing from three spatial dimensions to two spatial dimensions is that $L(z)$ is given by~\eqref{eqn:L-integral} but with $\reg$ being two dimensional and the integral over $S_2$ being a line integral.
    
    Now, to show that the discrete pair-correlation function \eqref{eqn:reduce-pair-corr-2} is equivalent to \eqref{eqn:dirac-radial} involving the Dirac delta, we need to show that:
    \begin{equation} \label{lim:counting to dirac}
      \lim_{\mathrm d z \to 0} \frac{\# \mathcal X^s_i(z)}{\mathrm d z} = \sum_{r_j \not = r_i, r_j \in \mathcal X^s} \delta(|\rv_i - \rv_j| - z).
    \end{equation} 
    To verify \eqref{lim:counting to dirac}, let us define:
    \[
    f(x) = \begin{cases}
      1, & \text{if} \quad z - \mathrm d z/2 \leq x \leq z + \mathrm d z/2,
      \\
      0, & \text{otherwise}.
    \end{cases}
    \] 
    Then, $\# \mathcal X^s_i(z) = \sum_{r_j \not = r_i, r_j \in \mathcal X^s} f(|\rv_i - \rv_j|)$. To complete the demonstration of \eqref{lim:counting to dirac} we note that for any $x$ we have that:
    \begin{equation} \label{eqn:equiv_expr_discr_cont}
      \lim_{\mathrm d z \to 0} \frac{f(x)}{\mathrm d z} = \delta(x - z).
    \end{equation}
    Equation \eqref{eqn:equiv_expr_discr_cont} becomes infinitely high and narrow while integrating to 1, precisely picking out values at $x=z$.

    \subsection{Particles in two regions} \label{sec:two region discrete}
    
    Using \eqref{p:two-regions} and the steps shown in \Cref{sec:two-regions}, we can reach a formula which is analogous to \eqref{eqn:reduce-pair-corr-2} but for $\rv_i \in \reg_1$ and $\rv_j \in \reg_2$ as illustrated in \Cref{fig:finite-sets} given by
     \begin{equation} \label{eqn:simple-pair-correlation}
      \pair_3(z) =  \frac{1}{S} \frac{1}{4\pi z^2 d z} \sum_{s}^S \frac{1}{\numdensity J_1}  \sum_{r_i \in \mathcal X^s \cap \reg_1} \# \mathcal X^s_i(z), \quad \text{(3D isotropic)},
    \end{equation}
    where we substituted \eqref{eqn:L-sphere} and used $J_1 = \# [\mathcal X^s \cap \reg_1]$. Equation \eqref{eqn:simple-pair-correlation} is the same as the formula \cite[Equation (2)]{barker1971monte}, and \cite[Equation (8.3.8)]{Tsang2001} when specialising their formula for particles with the same radius, and when taking $\reg_1 = \reg_2$. 
    
    For two spatial dimensions the pair-correlation becomes
    \begin{equation} \label{eqn:simple-pair-correlation-2D}
      \pair_2(z) =  \frac{1}{S} \frac{1}{2\pi z d z} \sum_{s}^S \frac{1}{\numdensity J_1}  \sum_{r_i \in \mathcal X^s \cap \reg_1} \# \mathcal X^s_i(z), \quad \text{(2D isotropic)}.
    \end{equation}
    
    As a final check, note that if particle positions were uncorrelated then $\# \mathcal X^s_i(z) \approx 4\pi \numdensity z^2 \mathrm dz$, for three spatial dimensions, because the number of particles would then be proportional to the volume times the number density. Substituting this approximation into \eqref{eqn:simple-pair-correlation} leads to $\pair_3(z) = 1$, which is what is expected from uncorrelated particles. 
   
    \section{The Structure Factor}\label{sec:structure factor}

    The structure factor, known as $\mathrm S(k)$, plays a crucial role in material science and condensed matter physics \cite{Pusztai1996, Pusztai1997, McGreevy1988, McGreevy2001, Stillinger2004}. Many experimental techniques, such as X-ray diffraction, neutron scattering and electron microscopy, inherently measure the structure factor, providing a deep understanding of the arrangement and density fluctuations of particles within a material \cite{Caballero2008, Micoulaut2022, Torquato2002}. Unlike the pair-correlation function $\pair (r)$ which gives the probability of finding particle pairs at a certain distance, the structure factor captures the intensity of scattered waves from materials, revealing both local and long-range order through its relationship as the Fourier transform of $\pair (r)$ \cite{Uche2006, Costin2004, Kuna2007}. This makes the structure factor, $\mathrm S(k)$, an integral part of interpreting experimental data, providing information for both characterising new materials and designing them with specific properties.
    
    The expressions for the structure factor in two and three dimensions are well-established in the field of material science \cite{Salvatore2002, Uche2006, Piekarewicz2012}. In this section we re-derive the structure factor from first principles using \eqref{eqn:isotropic-L-delta-3D} and \eqref{eqn:isotropic-L-delta-2D}.

    Mathematically, the pair-correlation $\pair(z)$ in the form \eqref{eqn:isotropic-L-delta-3D} shows how $\pair(z)$ is a discontinuous function when calculated from a finite number of particles. It is discontinuous in the variables $z$ and the position of the particles $\rv_i$. As $\pair(z)$ is not a smooth function, we can not use techniques from local nonlinear optimisation to calculate a particle configuration to match a specific pair-correlation. To avoid this, we can take a transform of the pair-correlation such as the structure factor: 
    \begin{align} \label{def:structure-factor}
      \mathrm S(k) = 1 + \numdensity \int (\pair(r) - 1) \eu^{-\iu \vec k \cdot \rv} \mathrm d \rv, \quad \text{(The Structure factor)},
    \end{align} 
    where $r = |\rv|$ and $k = |\vec k |$ is the magnitude of the wave vector. Equation \eqref{def:structure-factor} matches the typical definition of the structure factor \cite{vynck2021light,Tsang2001,Salvatore2002}, noting that the notation $h(r) = \pair(r) - 1$ is commonly used. 
    
    For some of the following calculations, we will perform the integral over $\pair(r)$ and $-1$ separately. To do this we note that:
    \begin{equation} \label{eqn:delta fourier}
      \delta(\vec k) =  \frac{1}{(2\pi)^n}\int \eu^{-\iu \vec k \cdot \rv} \mathrm d \rv, 
    \end{equation} 
    where $n$ is the spatial dimension, which in this paper is either $n = 3$ or $n=2$.
    
    For any given isotropic pair-correlation $\pair(r)$ the structure factor~\eqref{def:structure-factor} can be simplified to a 1D integral. To do this, we assume that particles become uncorrelated at a distance of $R$ so that $\pair(r) = 1$ for $0 \leq r \leq R$. Then from \eqref{def:structure-factor} we can calculate 
    that for three spatial dimensions:
    \begin{equation} \label{eqn:S3 numerical}
        \mathrm S_3(k) = 1 + \frac{4 \pi}{k} \numdensity \int_0^R (\pair_3(r) - 1) \sin (k r) r \mathrm d r,
    \end{equation}
    and for two spatial dimensions:
    \begin{equation} \label{eqn:S2 numerical}
        \mathrm S_2(k) = 1 + 2 \pi \numdensity \int_0^R (\pair_2(r) - 1) \mathrm J_0(k r)   r \mathrm d r.
    \end{equation}
    We will use \eqref{eqn:S2 numerical} to calculate the structure factor from a target pair-correlation in the next section.
    
    We can further simplify the structure factor when calculating it from a configuration of particles.  For three spatial dimensions, we can substitute \eqref{eqn:isotropic-L-delta-3D} into the structure factor~\eqref{eqn:S3 numerical}, but with $R = \infty$, and using the property of the Dirac delta, to obtain:
    \begin{equation} \label{eqn:structure-factor-3D}
      \tcboxmath{
        \mathrm S_3(k)  = 1 + \frac{1}{S} \sum_{s}^S \frac{1}{J_1}  \sum_{\rv_i\in \mathcal X^s \cap \reg_1}  \sum_{\underset{\rv_j \not = \rv_i}{\rv_j\in \mathcal X^s} } 
        \frac{\sin(k |\rv_i - \rv_j|)}{ k |\rv_i - \rv_j|},
      } \quad \text{(3D Structure factor),}
    \end{equation}
    for $k>0$. When calculating \eqref{eqn:structure-factor-3D}, a term of the form $- (2\pi)^3 \numdensity \delta(\vec k)$ appears, according to \eqref{eqn:delta fourier}, however, as we consider only $k>0$ it has no contribution to the above.
    
    Following analogous steps, the two dimensional structure factor calculated from \eqref{eqn:isotropic-L-delta-2D} becomes:
    \begin{equation}\label{eqn:structure-factor-2D}
      \tcboxmath{
      \mathrm S_2(k) = 1 +  \frac{1}{S J_1} \sum_{s}^S  \sum_{\rv_i\in \mathcal X^s \cap \reg_1}  \sum_{\underset{\rv_j \not = \rv_i}{\rv_j\in \mathcal X^s} } 
      \mathrm J_0(|k||\rv_i - \rv_j|),
      } \quad \text{(2D Structure factor),}
    \end{equation}
    for $k>0$, where $\mathrm J_0$ is the Bessel function of the first kind. 

    \section{Particle configurations from the structure factor}\label{sec:particles from structure}

    In theory, the pair-correlation is calculated by taking into account an infinite number of different particle configurations. Yet, many exotic material properties can be achieved by choosing specific pair-correlations, with one example being hyperuniform disordered materials \cite{Torquato2018, Salvatore2002,Zhang2020}. So being able to calculate one configuration of particles which closely represents any given pair-correlation would provide a route to fabricate particulate materials which exhibit exotic properties. 
    
    To recover a specific configuration of particles from a pair-correlation, we show a method to find a configuration of particles which is the mean particle configuration. Suppose we are given some pair-correlation $\mathrm g^\star(z)$, then we want one configuration of particles that when substituted into \eqref{eqn:simple-pair-correlation} will be close to $\mathrm g^\star(z)$, when removing the sum over the ensembles by setting $S=1$. 
    
    \subsection{Restrictions} \label{sec:restrictions}
    For any given pair-correlation $\pair$ and $\mathrm S$ structure factor there are certain restrictions \cite{Yamada1961,Costin2004,Uche2006} that need to be satisfied. These need to be considered when choosing a target pair-correlation $\pair^\star$ or structure factor $\mathrm S^\star$. We will only consider a few of these restrictions, and note that there may be an infinite number of necessary, though more complicated, conditions on the pair-correlation \cite{Costin2004}. 
    
    The simplest restrictions are that:
    \begin{equation}
      \pair^\star(r) \geq 0 \quad \text{and} \quad \mathrm S^\star(k) \geq 0,  
    \end{equation}
    where the first is due to the definition~\eqref{def:pair-correlation-two} together with the basic rule that probability functions must be positive. The second needs to be non-negative because of the relation of the structure factor $\mathrm S(k)$ with the variance of the particle density. This is discussed in \cite{Costin2004,Stillinger2004, Uche2006,Yamada1961}.
    
    As we focus on disordered particulates we require that particles become uncorrelated at some distance $R$. This requirement implies that:
    \begin{equation} \label{cond:disordered}
      \pair^\star(r) = 1, \qquad \text{for every} \;\; r \geq R.
    \end{equation}
    Further, from the definition~\eqref{def:pair-correlation-two}, together with \eqref{p:two-regions},  we have that: 
    \begin{align} \label{eqn:integrate pair}
      \int_{\reg_2}\int_{\reg_1} \pair(\vec x_1 , \vec x_2)  \mathrm d \vec x_1 \mathrm d \vec x_2 &= \frac{J_2 - 1}{J_2} |\reg_1| |\reg_2| \int_{\reg_1}\int_{\reg_2}  \mathrm{p}(\vec x_1,\vec x_2)  \mathrm d \vec x_1 \mathrm d \vec x_2 \nonumber \\
      &=  \frac{ J_2 -1 }{J_2} |\reg_1| |\reg_2|.
    \end{align}      
    Alternatively, using isotropy so that $\pair$ depends on only $|\vec x_1 - \vec x_2|$, the condition \eqref{cond:disordered}, and specialising to three spatial dimensions, we make use of \eqref{eqn:isotropic-L-delta-3D} to obtain:   
    \begin{align} \label{eqn:integral isotropic pair 3d}
      \int_{\reg_2}\int_{\reg_1} (\pair_3(|\vec x_1 - \vec x_2|) - 1) \mathrm d \vec x_1 \mathrm d \vec x_2 &= \int\int_{0}^R\int_{\reg_1} (\pair_3(z) - 1) \mathrm d \vec x_1 z^2 \mathrm d z \mathrm d \Omega \nonumber \\
      &= 4 \pi  |\reg_1|\int_{0}^R (\pair_3(z) - 1)  z^2 \mathrm d z.
    \end{align}
    In \eqref{eqn:integral isotropic pair 3d}, the integral over the regions $\reg_1$ and $\reg_2$ are simplified to a radial integral over distance $z$ with an angular component represented by $\Omega$, the solid angle. The solid angle $\Omega$ is crucial for integrating over all directions in three-dimensional space, accounting for the isotropy condition \eqref{cond:disordered}. By changing variables to $\vec z = \vec x_2 - \vec x_1$, we further simplify the expression, resulting in the integral over $z$ and $\Omega$ that accounts for the entire spherical symmetry around a point. Combining \eqref{eqn:integrate pair} and \eqref{eqn:integral isotropic pair 3d} leads to the restriction:
    \begin{equation}
     \int_{0}^R \pair_3(z)  z^2 \mathrm d z   =  \frac{ R^3}{3} -\frac{1 }{4 \pi \numdensity }, 
    \end{equation}
    where we used $\numdensity = J_2 / |\reg_2|$. For two spatial dimensions, following analogous steps and using \eqref{eqn:isotropic-L-delta-2D}, we obtain the restriction:
    \begin{equation} \label{eqn:integral isotropic pair 2d}
       \int_{0}^R \pair_2(z)   z \mathrm d z = \frac{R^2}{2}  -\frac{ 1 }{2 \pi \numdensity}.
    \end{equation}
    We can also translate the pair-correlation to satisfy \eqref{eqn:integral isotropic pair 2d}. Let:
    \begin{equation} \label{eqn:g_perturbation}
      \pair_2(r) = \pair_2^0(r) + \mathfrak{a} \ d p(r),
    \end{equation}
    where $dp(r) \to 0$ when $r \to R$. 
    In simple terms, \eqref{eqn:g_perturbation} modifies an initial pair-correlation function $\pair_2^0(r)$ to include a perturbation, aiming to adjust the particle distribution within a specific range. The perturbation $dp(r)$ is designed to become negligible as the distance $r$ reaches a specific point $R$. Then, given a number density $\numdensity$ we can obtain $dp$ from \eqref{eqn:integral isotropic pair 2d}:
    \begin{equation} \label{eqn:scaling_factor}
      \mathfrak{a}   = \Big [ \frac{R^2}{2}  -\frac{ 1 }{2 \pi \numdensity} - \int_{0}^R \pair_2(z) z \mathrm d z \Big ] \Big [ \int_{0}^R dp(z) z \mathrm d z \Big ]^{-1}.
    \end{equation}
    The selection of $dp(r) = \eu^{- 6r / R}$ guarantees this vanishing effect, exponentially decreasing as $r$ increases. The equation for $\mathfrak{a}$ calculates the necessary scaling factor to achieve the desired distribution pattern, factoring in the number density of particles and the integral of the pair-correlation function up to $R$.
    
    \subsection{Gradient optimisation}
    
    For some inner product:
    \[
      \langle \mathcal{G}, \mathcal{H} \rangle_k = \int \mathcal{G}(k) \mathcal{H}(k) w(k) \mathrm d k,  
    \]
    where $w(k)$ is some known weight. The objective is to find a particle configuration $\mathcal X$ that minimises:
    \begin{equation} \label{eqn:min f}
      \min_{\rv_i \in \mathcal X} f(\mathcal X), \quad \text{where} \quad f(\mathcal X) := \langle \mathrm S - \mathrm S^\star, \mathrm S - \mathrm S^\star \rangle_k.
    \end{equation}  
    Equation \eqref{eqn:min f} uses an integral to assess the similarity between the actual and target structure factors, $\mathrm S$ and $\mathrm S^\star$ respectively, over a range of spatial frequencies $k$. This method helps to quantify how much the particle configuration $\mathcal{X}$ deviates from the desired configuration.

    Most methods in the literature \cite{Manwart1999,Quiblier1984} achieve this by using non-gradient-based methods such as Genetic Algorithms, Nelder-Mead, and Simulated Annealing. However, $f$, as written above, is a smooth function of the position of the particles in $\mathcal X$. Further, it is straightforward to analytically calculate the gradient of $f$ in terms of the particle positions $\rv_j$. This implies the gradient based methods \cite{ruszczynski2011nonlinear} hold significant potential for superior performance over traditional non-gradient-based approaches. In specific, the objective function $f$, and its gradient, can be computational expensive to calculate so we opt to use the Limited memory BFGS (L-BFGS) methods \cite{liu1989limited, Mogensen2018}, as it stores information on the Hessian (the gradient of the gradient) and uses this to accelerate convergence. This often implies that L-BFGS requires less evaluations of the objective function and its gradient \cite{liu1989limited}.
    
    Specifically, we develop a method to minimise \eqref{eqn:min f} in two steps, one global, and one local. Separating the two steps allows us to search over a large area of the parameter space with the global step, while still obtaining high precision with the local step. Another clear reason, based on recovering a configuration of particles, is that if we had the constraint of particles not overlapping, then it can lead to particles being locked in configurations which can be far from the global minimum. For this reason we only enforce no particle overlapping in the local step.
    
    \paragraph{The global step.} A global optimisation that completely rearranges all particles to minimise~\eqref{eqn:min f}. For this step, we will allow particles to overlap, helping us explore a broad range of particle configurations and avoid locking the particles in a configuration and we will use a limited range for the wavenumbers $k_1 \leq k \leq k_2$ when minimising \eqref{eqn:min f}. That is, in this step we do not want to resolve spatial details smaller than the length scale $2a$, since it might not significantly influence the properties of the material. So the shortest wavelength $\lambda$ we consider for the structure factor is $\lambda = 2a$ which corresponds to wavenumbers $k \leq k_2 = \pi / a$. The smallest wavenumber $k_1$ is determined by the dimensions of the material: let $D_1$ be the smallest dimension of $\reg_1$, then the longest wavelength we consider is $\lambda = D_1$ which implies that $k \geq k_1 = 2\pi / D_1$.   
    
    \paragraph{The local step.} This step improves the particle configurations obtained from the global step. This involves making small adjustments to the particle positions to achieve a more accurate and realistic distribution of particles. We also enforce a penaliser $W$, shown in \eqref{eqn:penaliser}, to prevent overlaps between particles. The penaliser has the form of an exponential that increases very rapidly as the distance decreases. This means that as particles get closer to each other, the penalty for their overlap increases exponentially. For this step we also want to resolve spatial details. Suppose we want to resolve details up to $a / 4$, then $k_2 = 8\pi$ and $k_1 \leq k \leq k_2$ for this step. 
    
    To use techniques from nonlinear optimisation~\cite{ruszczynski2011nonlinear, Mogensen2018} to minimise calculate \eqref{eqn:min f} we need to calculate the gradient:
    \begin{align}
     \frac{\partial f(\mathcal X)}{\partial \rv_j} =  \frac{\partial }{\partial \rv_j} \langle \mathrm S - \mathrm S^\star, \mathrm S - \mathrm S^\star \rangle_k  
    =  2\sum_i \langle \frac{\partial \mathrm S}{\partial R_{ij}}, \mathrm S - \mathrm S^\star \rangle _k \frac{\partial R_{ji}}{\partial \rv_j}  
    \\
    = 2  \langle \sum_i \frac{\partial \mathrm S}{\partial R_{ji}} \frac{\vec R_{ji}}{R_{ji}}, \mathrm S - \mathrm S^\star \rangle_k,
    \end{align}
    where
    \begin{equation}
      \begin{cases}
      \vec R_{ji} = \rv_j - \rv_i, & \text{the vector from } \rv_i \text{ to } \rv_j, \\
      R_{ji} = |\rv_j - \rv_i|, & \text{the magnitude of } \vec R_{ji}, \\
      \dfrac{\partial R_{ji}}{\partial \rv_j} = \dfrac{\vec R_{ji}}{R_{ji}}, & \text{the gradient of } R_{ji} \text{ with respect to } \rv_j.
      \end{cases}
    \end{equation}
    In particular, for~\eqref{eqn:structure-factor-2D} we have:
    \begin{equation}
      \frac{\partial \mathrm S_2}{\partial R_{ji}} =  \frac{k}{J_1}
      \begin{cases}
         \mathrm J_0'(|k| R_{ji}), \quad & \text{if} \quad \rv_j \not \in \reg_1,
        \\   
         2 \mathrm J_0'(|k| R_{ji}), \quad & \text{if} \quad \rv_j \in \reg_1,
      \end{cases}
    \end{equation}
    where there are two cases because: if $\rv_j \in \reg_1$ then $\mathrm J_0(k R_{ji})$ gets summed twice in \eqref{eqn:structure-factor-2D}, but if $\rv_j \not \in \reg_1$ then $\mathrm J_0(k R_{ij})$ only appears once in the summation. Likewise for~\eqref{eqn:structure-factor-3D} we have:
    \begin{equation}
      \frac{\partial \mathrm S_3}{\partial R_{ji}} =  \frac{k}{J_1} \frac{1}{(k R_{ij})^2}
      \begin{cases}
        k R_{ij} \cos(k R_{ij}) - \sin(k R_{ij}), \quad & \text{if} \quad \rv_j \not \in \reg_1,
        \\   
        2 k R_{ij} \cos(k R_{ij}) - 2\sin(k R_{ij}), \quad & \text{if} \quad \rv_j \in \reg_1.
      \end{cases}
    \end{equation} 
    For most optimisation methods, we choose to use Optim.jl \cite{Mogensen2018}, we need to supply the total gradient:
    \[
    \nabla f = \left[\frac{\partial f(\mathcal X)}{\partial \rv_1}, \frac{\partial f(\mathcal X)}{\partial \rv_2}, \ldots, \frac{\partial f(\mathcal X)}{\partial \rv_j} \right],  
    \]
    where the block vector on the right is typically flattened to be just one large vector. \\
    For the local step, after the global step is complete, we add a restriction that penalises particles that are overlapping. That is instead of minimising \eqref{eqn:min f}, we minimise:
    \begin{equation} \label{eqn:min f and W}
      \min_{\rv_i \in \mathcal X} f(\mathcal X) + A W(\mathcal X),
    \end{equation}
    where $A$ is some large positive constant that is problem dependant and:
    \begin{equation} \label{eqn:penaliser}
      W = \sum_{i, j \not = i} \chi_{\{R_{ji} < 2a\}} \eu^{ - 4R_{ji}^2 / (2a)^2 }.
    \end{equation}
    The specific formula for the penaliser \eqref{eqn:penaliser} has the form of a Gaussian distribution \cite{kaipio2006statistical} and was chosen to effectively prevent particle overlaps during the local optimisation step. This function is smooth and differentiable, and it rapidly increases the penalty as the distance between particles becomes less than twice their radius. Having said that, the gradient of the penaliser \eqref{eqn:penaliser} with respect to the position of the $j$-th particle has the following form:
    \begin{equation}
      \frac{\partial W}{\partial \rv_j} =   - \frac{4}{a^2} \sum_{i \not = j} \chi_{\{R_{ji} < 2a\}}   \eu^{ - 4R_{ji}^2 / (2a)^2 }  \vec R_{ji}.
    \end{equation}
    For ease of implementation we use a discrete form:
    \[
    \langle \mathcal{G} , \mathcal{H} \rangle_k = \sum_q \mathcal{G}_q \mathcal{H}_q w_q,
    \]
    where $w_q$ are some Gaussian quadrature weights. For example we can write:
    \begin{equation} \label{eqn:objective_discrete_form}
        \frac{\partial f (\mathcal X) }{\partial \rv_j}  =  2  \sum_{i q} \frac{\partial \mathrm S_q}{\partial R_{ji}} \frac{\vec R_{ji}}{R_{ji}} (\mathrm S_q - \mathrm S^\star_q) w_q.
    \end{equation}
    The discrete form \eqref{eqn:objective_discrete_form} simplifies the calculation of gradients in computational simulations. This technique allows for efficient numerical approximation of gradients necessary for optimising particle positions based on the difference between observed and desired structural properties.

    \subsection{Preliminary numerical results} \label{sec:preliminary-results}
    
    In this section we share our preliminary results and discuss potential future developments with our method.

\begin{figure}[ht] 
      \centering
      \begin{tikzpicture}
        \centering
        \draw (-6.5,0) node {\includegraphics[width=0.45\textwidth]{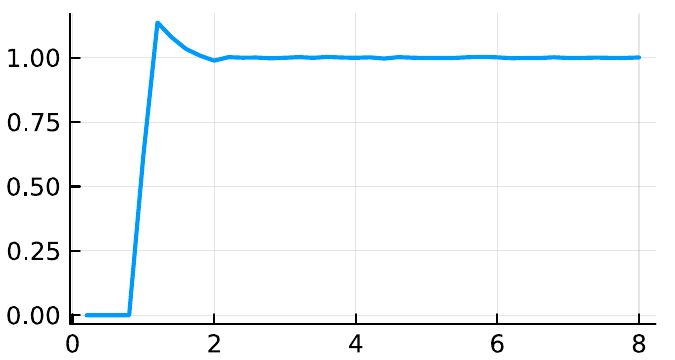}};
        \node[rotate=90] at (-10.65,0.1) {Pair-correlation};
        \node at (-6.3,-2.4) {z (distance)};
        \node at (-6.3,2.4) {\large Percus-Yevick};
        \draw (1.3,0) node {\includegraphics[width=0.45\textwidth]{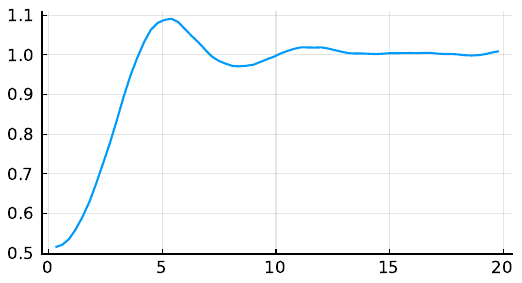} } ;
        \node[rotate=90] at (-2.75,0.1) {Structure factor};
        \node at (1.73,-2.4) {k (wavenumber)};
        \node at (1.73,2.4) {\large Percus-Yevick};
    \end{tikzpicture}
      \caption{On the left, the Percus-Yevick pair-correlation for hard discs \cite{adda2008solution}, in two spatial dimensions, where the discs occupy 15\% of the volume fraction. On the right the corresponding structure factor when using~\eqref{eqn:S2 numerical}.}
      \label{fig:Percus-Yevick-fig}
    \end{figure}
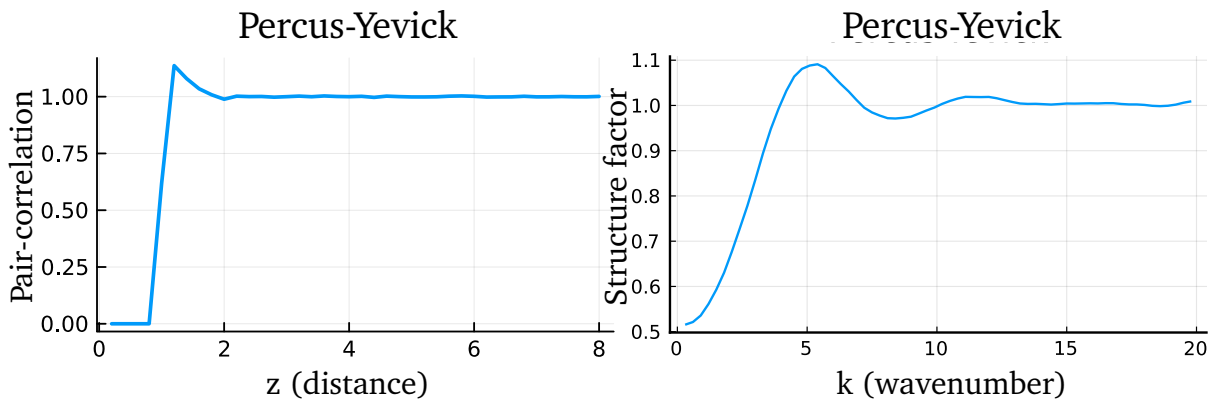

\begin{figure}[H]
      \centering
      \begin{subfigure}{0.49\linewidth}
          \centering
          \includegraphics[width=0.6\linewidth]{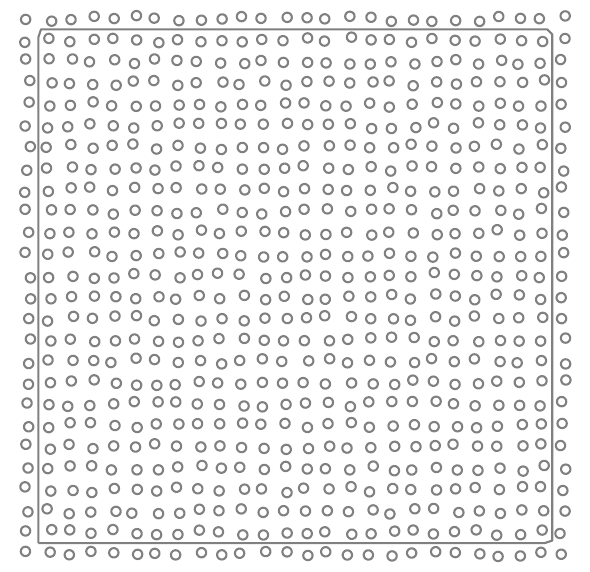}
          \caption{Initial particle configuration.}
          \label{fig:initial_config}
      \end{subfigure}
      \begin{subfigure}{0.49\linewidth}
          \centering
          \includegraphics[width=0.6\linewidth]{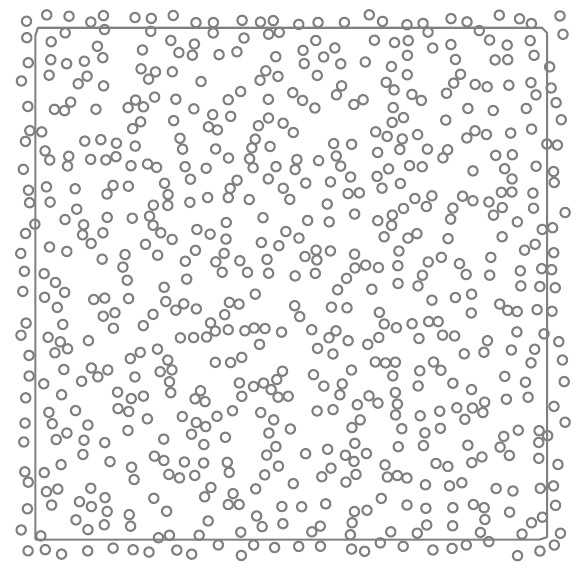}
          \caption{Predicted particle configuration.}
          \label{fig:predicted_config}
      \end{subfigure}
      \caption{\Cref{fig:initial_config} presents the initial position of all the particles, while \Cref{fig:predicted_config} demonstrates the optimised particle configuration that closely aligns with the structure factor depicted in \Cref{fig:Percus-Yevick-fig}, achieved through our optimisation method.}
      \label{fig:configurations}
  \end{figure}

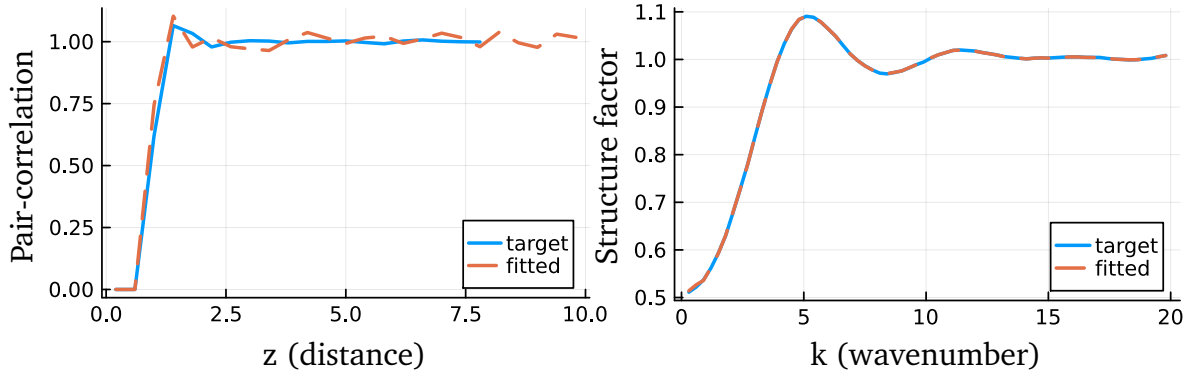
\begin{figure}[ht]
    \centering
    \begin{tikzpicture}
        \centering
        \draw (-6.5,0) node {\includegraphics[width=0.43\textwidth]{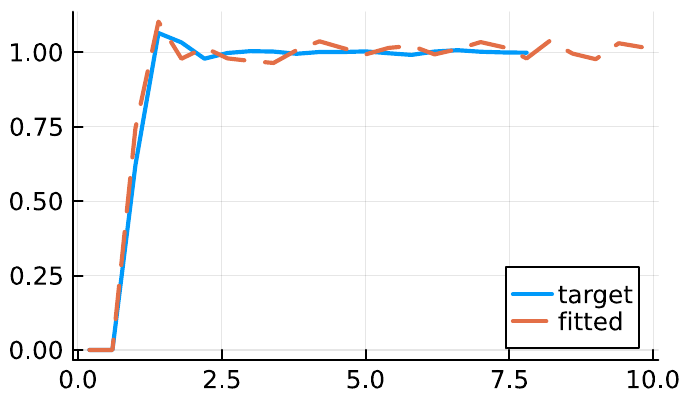} } ;
        \node[rotate=90] at (-10.55,0.1) {Pair-correlation};
        \node at (-6.25,-2.55) {z (distance)};
        \draw (1.3,0) node {\includegraphics[width=0.43\textwidth]{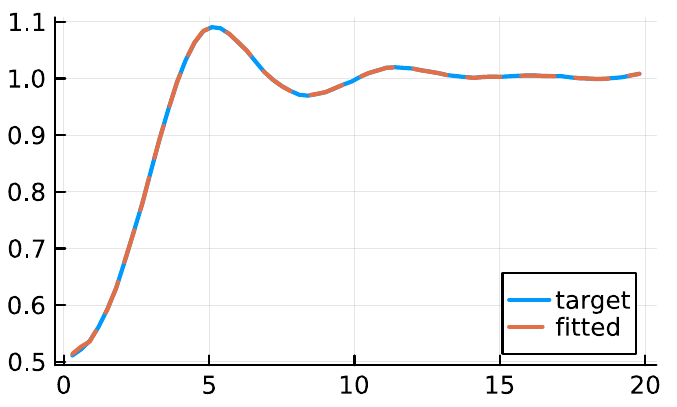} } ;
        \node[rotate=90] at (-2.78,0.1) {Structure factor};
        \node at (1.48,-2.55) {k (wavenumber)};
    \end{tikzpicture}
    \caption{The optimised particle configuration consisting of 600 particles of radius $a=1$ occupying 15\% of the particulate, depicted in \Cref{fig:predicted_config}, closely aligns with the desired structure factor and pair-correlation, despite noise introduced by the limited number of particles.}
    \label{fig:predict pair}
\end{figure}

    \paragraph{Selection of pair-correlation.} The first step is to have a systematic way to choose candidate pair-correlations, with one motivation being to control wave propagation \cite{gower2021effective}. Our work begins with the selection of appropriate pair-correlations that satisfy the restrictions given in \Cref{sec:restrictions}. Our choice is the Percus-Yevick model, a well-studied pair-correlation \cite{Tsang2001,Salvatore2002, adda2008solution, Gerhard2021} which represents disordered particles with very short range correlation. This model effectively catches the behaviour of uniformly distributed particles that exhibit correlations only because they can not overlap. Figure \ref{fig:Percus-Yevick-fig} illustrates this concept, displaying the pair-correlation for hard discs in a two-dimensional setup, where these discs account for a 15\% volume fraction and each particle has a radius of $a=1$. This figure also includes the corresponding structure factor, highlighting how the spatial arrangement of particles influences wave propagation through the material.
    \paragraph{Initial configuration.} The next step is to generate an initial configuration of particles with the correct sizes, and within a given volume fraction \cite{Tsang2001}, as shown in \Cref{fig:initial_config}. The simplest way to do this is to place particles on a grid. This also facilitates defining the regions $\reg_1$ and $\reg_2$ which are needed to calculate the pair-correlation without introducing artefacts from the boundary, as discussed in \Cref{sec:two-regions}. However, placing particles exactly in a periodic grid would lead to a set of problems, especially when it comes to finding the optimal arrangement using gradient-based optimisation techniques. The symmetrical nature of the periodic grid tends to position the configuration at a local maximum which may not be an ideal starting position for gradient-based methods. In other words, this symmetry can trick our optimisation methods into thinking they have found the best arrangement when there might be better choices they have not explored yet. To avoid this, we introduce a simple yet effective strategy: we slightly move each particle by a small distance in a random direction to break the symmetry, creating a more favourable setting for optimisation.
  \paragraph{Optimisation method and results.} Our optimisation method is executed in two steps: The first step of our method, the global step, minimises the objective function \eqref{eqn:min f} and is able to exactly match the specified structure factor, as demonstrated in \Cref{fig:Percus-Yevick-fig}. Following this global step, we employ a local optimisation step to refine the particle configuration further. The result of the structure factor of the optimised particle configuration, after the local step, is displayed in \Cref{fig:predict pair}. Despite the slight noise introduced by the finite number of particles (600 particles of radius $a=1$ occupying a 15\% volume fraction in two spatial dimensions), the predicted structure factor matches the target structure factor. Moreover, the predicted pair-correlation, shown in \Cref{fig:predict pair}, offers further insight. Although there is some noise due to the small number of particles, we see a good match with the desired pair-correlation.

  \section{Conclusions} \label{sec:discussion}
  
  In this paper, we deduced from first principles how to calculate both the pair-correlation and structure factor of a finite disordered particulate. We demonstrated how to do this without getting artefacts from the boundary of the particulate, something which seems to be ignored in the literature. Thus, it was possible to calculate the pair-correlation of an infinite particulate from a finite sample. This is generally desired, as most theoretical methods use the pair-correlations from an infinite medium, including the very important translation invariance which greatly simplifies the pair-correlation.
  
  Being able to calculate the pair-correlation from a particulate was rather straightforward in comparison to the inverse: calculating a particulate from a pair-correlation. We presented a method to calculate one configuration of particles that best fitted a given pair-correlation or structure factor. 
  
  Most methods in the literature \cite{Uche2006,Horowitz2004,ruszczynski2011nonlinear} that calculate particle configurations from the pair-correlation of structure factor use non-gradient-based methods such as Genetic Algorithms, Nelder-Mead, and Simulated Annealing. Several works in fact focus on trying to approximate the pair-correlation with a particulate, which is an inherently discontinuous function as the joint probability function~\eqref{eqn:p-joint-dirac} involves Dirac deltas. Instead, we suggested that it is best to seek a configuration of particles to approximate the structure factor, as it is a smooth function of the particle positions, as shown by \eqref{eqn:structure-factor-3D} and \eqref{eqn:structure-factor-2D}. This enabled us to analytically calculate the gradient of the structure factor, in terms of the particle positions, leading to faster convergence to optimal configurations. The efficacy of our method was evidenced through visual representations in Figure \ref{fig:configurations}, which showed the transformation from the initial to the optimised particle configurations, matching the theoretical model. Furthermore, the comparison of the structure factors before and after optimisation, as depicted in Figure \ref{fig:predict pair}, underscored the precision of our approach, achieving an almost identical match to the targeted structure factor. 
  
  Optimising particle configurations holds significant implications for a wide range of engineering applications. By precisely controlling particle configurations, engineers can tailor material properties for diverse applications in aerospace, automotive, and structural engineering, ensuring optimal performance and reliability.

  \paragraph{Future avenues.} We presented a two step method to calculate the structure factor from a configuration of particles, one step for global optimisation that avoids particle locking, and one step for local optimisation. To have clear evidence about the performance advantage of our method, it is essential to further validate and compare our method against traditional brute-force, non-gradient-based optimisation techniques, such as Genetic Algorithms and Simulated Annealing. This method can be further developed to easily add priors about the particle configuration and treat this as a statistical inverse problem \cite{kaipio2006statistical}. There is a significant amount of prior information that could be used \cite{Patelli2009,Yeong1998}. For example, when using more than one type of particle, some particles may repel or attract each other. Or there can be specific knowledge on chains or sub-components of particles. This information can be added as a prior, or regulariser. This seems to be an unexplored approach that could greatly increase the performance of this inverse problem. Moreover, the preliminary results presented in \Cref{sec:preliminary-results}, not only validate our novel approach to determining particle configurations from structure factors, but also open up new possibilities for future research. By fine-tuning the particle configurations, we could explore a wider range of pair-correlations and structure factors, potentially uncovering new ways to manipulate wave propagation in disordered particulate materials.

  \section*{Author contributions}
  AK conceived of the study, drafted the manuscript, developed the theoretical calculations, wrote all the code for the numerical calculations, and produced all the figures. ALG helped conceive the study, edited the manuscript, assisted with and verified the theoretical calculations.
  
  \section*{Data and reproducibility}
  \label{sec: data_chapter4}
  
  To produce our results we used the open source software \cite{2020MultipleScatering.jl}, \cite{2023ParticleCorrelations.jl} and \cite{Optim.jl-2018}.

  \newpage
  
  \printbibliography[heading=subbibintoc]
    
\end{refsection}

\end{document}